\documentclass[pra,a4paper,twocolumn,superscriptaddress]{revtex4}

\usepackage{latexsym}
\usepackage{amsmath}
\usepackage{amssymb}
\usepackage{amsfonts}
\usepackage{comment}
\usepackage{bbm}
\usepackage{color}

\newcommand{\dif}{\mbox{d}}
\newcommand{\tr}{{\textrm{tr}}}

      \usepackage{graphics}

\begin{document}

\title{Macroscopic Entanglement and Phase Transitions\\[0.6cm]
\normalsize talk given at the XXXVIII Symposium on Mathematical Physics\\ 
\emph{Quantum Entanglement \& Geometry}\\
 in Toru\'n, June 4-7, 2006}

\author{Janet Anders\footnote{email: janet@qipc.org}}
\affiliation{Quantum Information Technology Lab, Department of Physics, National University of Singapore, Singapore 117542}
\author{Vlatko Vedral}
\affiliation{The School of Physics and Astronomy, University of
Leeds, Leeds LS2 9JT, UK} 

\date{\today}

\maketitle

\textbf{Abstract}
This paper summarises the results of our research on macroscopic entanglement in spin systems and free Bosonic gases. We explain how entanglement can be observed using entanglement witnesses which are themselves constructed within the framework of thermodynamics and thus macroscopic observables.  These thermodynamical entanglement witnesses result in bounds on macroscopic parameters of the system, such as the temperature, the energy or the susceptibility, below which entanglement must be present. The derived bounds indicate a relationship between the occurrence of entanglement and the establishment of order, possibly resulting in phase transition phenomena. We give a short overview over the concepts developed in condensed matter physics to capture the characteristics of phase transitions in particular in terms of order and correlation functions.
Finally we want to ask and speculate whether entanglement could be a generalised order concept by itself, relevant in (quantum induced) phase transitions such as BEC, and that taking this view may help us to understand the underlying process of high-T superconductivity.

\section{Introduction}

\emph{Entanglement} is a purely quantum mechanical feature which has been studied extensively in the last 20 years. For a long time people would regard the property of being entangled as timid, hidden and easily destroyed. However, in recent research we have learned that entanglement naturally exists on a macroscopic scale in quantum many-body systems, e.g. spin chains, gases, crystals, and it will be exhibited even in the thermodynamical limit and even at finite temperatures. This comes as a surprise since thermodynamical systems , being so big, seem quite classical, whereas the very nature of entanglement is quantum. Here we want to show that the quantum properties, in particular entanglement, are related to the known thermodynamical properties in macroscopic systems.

In many-body systems, the exact microscopic configuration of all particles contains in fact much more information ($6 \times 10^{23}$) than we usually need to understand the system, its capabilities and its response to external changes.  Thermodynamics is the theory which tells us how to describe such systems by assigning \emph{macroscopic properties} alone, for instance, it defines state variables, potentials and response functions. Using the language of thermodynamics, we want to know, under what conditions, e.g. temperature $T$, magnetic field $B$, particle density $\rho$, ...,  is entanglement present? And how can we detect  and extract (see References \cite{06Heaney, ent-extract}) such macroscopic entanglement? And from a practical viewpoint: Can we use macroscopic entanglement  for the construction of a quantum computer? And finally, does macroscopic entanglement reveal something new about the occurrence of phase transitions in quantum systems and could it possibly be one ingredient of a new theory of high-$T_c$ superconductors? 

Addressing entanglement as a thermodynamical property has become a successful exercise. It was first mentioned in \cite{A} and many publications discussing various aspects of entanglement in macroscopic systems followed, for instance \cite{B}. In the present paper we summarise the results of our investigations in the field of macroscopic entanglement. The outline of the paper is as follows. In Section \ref{sec:th+ew}, we review the key ingredients of our work - thermal states and entanglement witnesses. 
	
	In Section \ref{sec:III} we discuss three examples where thermodynamical state variables can serve as macroscopic entanglement witnesses. 
	Firstly, the Heisenberg model, discussed in references \cite{Arnesen, Brukner}, for which a combination of thermodynamical quantities is identified as an entanglement witness leading to a critical parameter range for the temperature $T$ and the external magnetic field $B$. 
	Secondly, we review the results of \cite{06Brukner} about the entanglement properties of the compound  [Cu(NO$_3$)$_2$ 2.5 D$_2$O], which can be approximated by the Heisenberg model for an alternating spin chain. In this example the magnetic susceptibility is identified as an entanglement witness and using measurement results from 1963  \cite{Berger}, it can be concluded that indeed the compound powder will become macroscopically entangled when cooled below approximately 5K. 
	The last example is a free, non-interacting Bosonic gas discussed in \cite{06Anders}. Reviewing this work, we will embark on a short discussion of what spatial entanglement in such a second quantised and continuous system can refer to, and find that the energy can serve as an entanglement witness for spatial entanglement between regions in space. We will also see, that the transition temperature for the occurrence of such entanglement is almost identical to the BEC critical temperature in 3D.
	
	Keeping this observation in mind,  we proceed to Section \ref{sec:PTs}, where  we give an overview of various phase transitions (PTs) and how they are characterised in condensed matter physics. In particular, we discuss criticality indicators such as the order-parameter appearing when symmetry-breaking occurs and concepts of \emph{order} such as long-range order (LRO), off-diagonal-long-range order (ODLRO), quasi-order. 
	
	In Section \ref{sec:MEO} we speculate that macroscopic entanglement may be a more general, all-round order parameter indicating the transition of a disordered phase towards an ordered phase;  an entanglement-ordered phase. We give an outlook on why this may prove useful in the search for a new theory of high-$T_c$ superconductivity. In Section  \ref{summary} we give possible future directions and  summarise the key points of this paper.
	
\section{Thermal states and entanglement witnesses}\label{sec:th+ew}

The statistical treatment of macroscopic systems in thermal equilibrium is summarised by the notion of the partition function in classical physics and similarly in quantum physics by the  \emph{thermal state} $\rho_T$, which carries all the macroscopic information of the system, regardless of microscopic configurations. The thermal state is determined by the Hamilton operator $\hat H$ and the temperature $T$, as the operator-valued quantity
\begin{equation}
	\rho_T = { 1 \over Z} \, e^{-{\hat H / k_B T}},
\end{equation}
where $Z = \tr[e^{-{\hat H /k_B T}}]$ is the quantum version of the partition function.  This partition function defines the  probability distribution for the occurrence of the ground state $|e_0 \rangle$ and the excited states $|e_j \rangle, j =1, 2, ...$, and allows us in principle to deduce any ensemble average we are interested in. 

Generally speaking, we want to find out whether a system in a thermal state is entangled for a given temperature and a specified partition, e.g. one half of a spin chain with respect to the other half. (When no partitioning is explicitly  specified, the term ``entangled'' is identical to ``not fully separable'' between the constituent subsystems.)   The thermal state expanded in the energy eigenbasis is a mixture,
\begin{equation}\nonumber
	\rho_T  = {1\over  Z} \left(  e^{-{E_0 / k_B T}} |e_0 \rangle\langle e_0| + e^{-{E_1 / k_B T}} |e_1 \rangle\langle e_1| + ...\right),
\end{equation} 
of the ground state $|e_0\rangle$, which is often entangled, with (possibly entangled) excited states $|e_j \rangle$, $j>0$, having energies $E_j > E_0$. To begin with, it is exactly this mixing of all states, that  makes it tough to calculate the elements of the density matrix explicitly. And even if we could do that, there are no good entanglement criteria for complex systems at hand, enabling us to decide whether the  actual density matrix is in fact entangled.

\begin{figure}[t]
  	\begin{center}
   	\scalebox{0.6}{\includegraphics{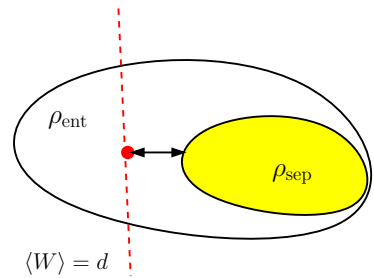}}
    	\caption{\label{fig:EW} The figure shows the set of separable states $\rho_{\rm{sep}}$ (shaded) and the set of entangled states $\rho_{\rm{ent}}$. When an observable $W$ of the system is measured (many times), and an expectation value $d$ is obtained, the set of possible states corresponding to that value is a hyperplane inside the set of states, here represented by the dashed line. If the hyperplane contains purely entangled states we have ``witnessed'' entanglement, without knowing what exactly the state is. The expectation value $d$ is also a related to the amount of entanglement, since many measures of entanglement (e.g. the relative entropy of entanglement) can be phrased in terms of the distance to the set of separable states. }
       	\end{center}
\end{figure}

A different approach is to measure an observable which distinguishes two sets of states - a set containing purely entangled states and a set containing separable and possibly entangled states. Such observables are called \emph{entanglement witnesses} (EWs), \cite{EWs},  because their measurement outcomes might \emph{witness} the presence of entanglement, see Figure \ref{fig:EW}. 
Using the smart concept of entanglement witnesses our previous quest to decide whether a thermal state is entangled simplifies to the question: Can entanglement be \emph{witnessed} by taking ensemble averages?  The answer is: Yes,  you can!, and we will discuss three particular systems to support our claim. For a discussion of  this question for  general thermal states please refer to \cite{Markham},  for instance.

\section{Thermodynamical variables as entanglement witnesses}\label{sec:III}

\subsection{Heisenberg model}

One of the first papers discussing entanglement occurring naturally in thermal systems is the paper by Arnesen {\it et al.} \cite{Arnesen}. As an example they take a simple dimer, made of two spin-${1 \over 2}$ particles coupled via the Heisenberg Hamiltonian $\hat H = - J  \vec \sigma_1 \cdot \vec \sigma_2$, with $J > 0$ for the anti-ferromagnetic case. They diagonalise the Hamiltonian and calculate an exact expression for the concurrence (an entanglement measure). With this expression, the authors find, that for zero magnetic field, when the singlet is ground state, the entanglement is maximal while for increasing $T$, the entanglement decreases due to the add-mixing of the excited states. But there is also the case when the temperature remains at absolute zero, and an additional external magnetic field $B$ is turned on. In this case the entanglement also decreases and there exists a critical value $B_c = 4J$ at which the entanglement vanishes completely since $|00\rangle$ becomes the ground state. This is a simple example of a quantum phase transition \cite{QPT} where the system changes abruptly its ground state. The authors extended their discussion also to spin chains of $N$ spins (which they solve numerically) and find qualitatively similar behaviour of the entanglement with regard to the parameters $T$ and $B$. 

The Heisen\-berg spin chain was re\-vised again by Brukner {\it et al.} \cite{Brukner} in 2004. Instead of a numerical investigation the authors here use the concept of entanglement witnesses and identify the total energy, $U + B M$, as an appropriate operator distinguishing between separable and entangled states.  For the Heisenberg XX or XXX model with nearest neighbour coupling $J$ and Hamiltonian $\hat H = - J \sum_j \vec \sigma_{j} \cdot \vec \sigma_{j+1}$, the expectation value $\langle \vec \sigma_i \cdot \vec \sigma_k \rangle$ in a fully separable configuration must factorise, and using a Cauchy-Schwartz inequality, the bound on the total energy is
\begin{equation}\label{eq:U+BM}
	\left|{U +B M} \right |\le N |J|,
\end{equation}
where $U$ is the internal energy, $B$ the external magnetic field, $M$ the magnetisation of the chain and $N$ the number of spins. If the expectation values $U$ and $M$ of a particular state together with the external field $B$ are such, that this inequality is violated, the state must be entangled. 
\begin{figure}[t]
  \begin{center}
   \scalebox{0.4}{\includegraphics{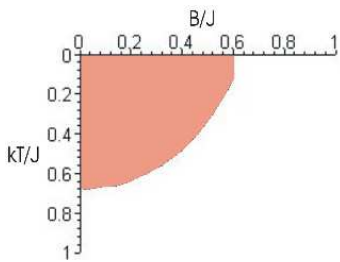}}
    \caption{\label{fig:Brukner} For a given coupling strength $J$, the thermal state of a Heisenberg spin chain is defined by the temperature $T$ and the external magnetic field $B$. When the values of $T$ and $B$ lie within the shaded area, i.e. $	\left|{U +B M \over NJ}\right| > 1$ , the state must be entangled. This figure is taken from \cite{Brukner}.}
    \end{center}
\end{figure}

Unfortunately it is not easy to measure the internal energy directly. One can get $U$ indirectly by measuring the heat capacity for some temperature range and integrating over $T$; Or one can use thermodynamical relations which connect the internal energy and magnetisation to the state variables $T$ and $B$. These analytical formulae have been derived by Katsura \cite{Katsura} in 1962, and Brukner {\it et al.} use them to determine lower bounds on the critical values of the temperature and the magnetic field, below which entanglement must certainly occur. Figure \ref{fig:Brukner} is a reprint (taken from \cite{Brukner}) showing the region in which entanglement is detected by the witness Eq. \eqref{eq:U+BM}. So indeed, it is possible to detect entanglement using solely thermodynamical properties of the system. It is obvious now, that changing the state variables $T$ and $B$ will result in a change of the entanglement of the state of the system and we could possibly turn it on or off, or tune it to our convenience.

\subsection{Alternating spin chain}

\begin{figure}[b]
  \begin{center}
   \scalebox{0.45}{\includegraphics{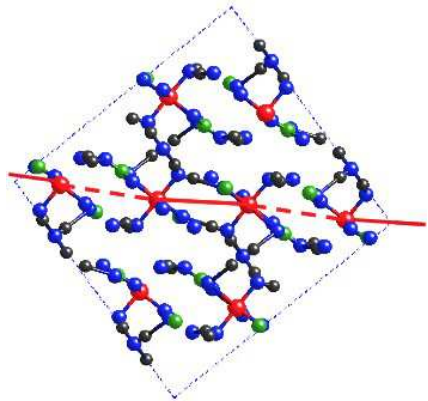}}
    \caption{\label{fig:cuprite} The picture shows the structure of the [Cu(NO$_3$)$_2$ 2.5 D$_2$O] molecules. The central copper atoms are displayed in red colour and the alternating binding forces  between the copper atoms are drawn with straight lines. The stronger coupling $J_1$ (solid line) occurs inside the molecule and effectively creates a dimer. The coupling which binds the established dimers to each other is much weaker, $J_2 << J_1$, and shown with dashed lines.}
       \end{center}
 \end{figure}

The dimer and the Heisenberg chain model can be used to describe some materials relevant in condensed matter physics in good approximation. In \cite{06Brukner}  Brukner {\it et al.} discuss the macroscopic entanglement present in a particular copper nitrate, [Cu(NO$_3$)$_2$ 2.5 D$_2$O]. The effective coupling between the central copper atoms can be approximated by, firstly, a stronger coupling inside the compounds with  coupling strength $J_1 \approx 0.44$meV which creates the dimers. Secondly, these dimers are arranged in a one-dimensional chain with nearest neighbour coupling, $J_2 \approx 0.11$meV, which is much  weaker  than the intra-dimer coupling, see Figure \ref{fig:cuprite}.

The Hamiltonian governing such an alternating Heisenberg chain of molecules is just 
\begin{equation}
	\hat H = - \sum_j \left( J_1 \vec \sigma_{2j} \cdot \vec \sigma_{2j+1} 
	+ J_2 \vec \sigma_{2j+1} \cdot \vec \sigma_{2j+2} \right),
\end{equation}
generating the thermal states $\rho_T = {1 \over Z} e^{- \hat H/k_B T}$.
The ground state, $\rho_{T=0}$, is an highly  entangled state and due to the alternating sequence of coupling constants  the system has an energy gap of  the order $\Delta \approx J_1$ between the ground and the first excited state. That means that as long as the temperature is low enough no excited state can be reached and the chain remains entangled. This simple argument gives an estimate for the transition temperature, $T_{trans} \approx J_1/k_B \approx 5$K, above which the entanglement of the ground state may be killed by the admixing of excited states. 

\begin{figure}[t]
  \begin{center}
   \scalebox{0.42}{\includegraphics{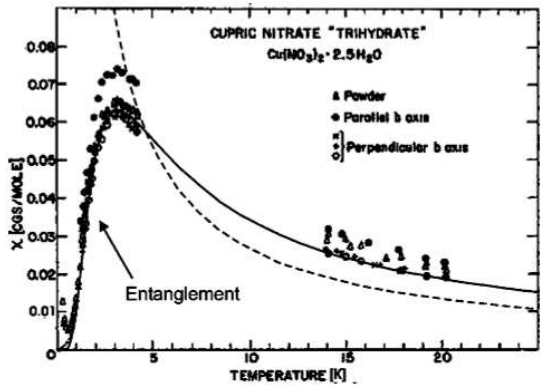}}
    \caption{\label{fig:suscep} The original figure showing measurement results (dots) of the magnetic susceptibility and the theoretical curve for $\chi (T)$ (solid line) is taken from \cite{Berger}. The additional transition curve $\chi_{trans} (T)$ (dashed line) and the indication of the entangled region was added by the authors of \cite{06Brukner}. The predicted $\chi (T)$-curve and the entanglement witness $\chi_{trans} (T)$ cut at $T=5$K. Measurement points below this temperature have all a lower susceptibility than any separable configuration could have. They must have been entangled in 1963!}
       \end{center}
 \end{figure}

In the paper \cite{06Brukner} the researchers identified the magnetic susceptibility as an entanglement witness and found the threshold value  
\begin{equation}
	\chi_{trans} = {1 \over 6} {g^2 \mu^2_B N\over k_B T}
\end{equation}
at the point where separability turns into entanglement. Here $g$ is the Land\'e-factor, $\mu_B$ Bohr's magneton and $N$ the number of compounds. Below this transition susceptibility $\chi_{trans}$ the observed system must necessarily be entangled. This threshold was then compared to experimental measurement results of the magnetic susceptibility of  [Cu(NO$_3$)$_2$ 2.5 D$_2$O]  which were published in \cite{Berger} in 1963. Astonishingly, there are a lot of measurement points in the region where entanglement must necessarily occur, see Figure \ref{fig:suscep}. This confirms nicely the claim that entanglement can exist at reasonably high temperatures and is macroscopically verifiable. Additionally the theoretical value of $\chi$ crosses the threshold value $\chi_{trans}$ at roughly 5K in great agreement with the heuristic argument above. So, the signatures of entanglement have been there all along - without people knowing it at that time!

\subsection{Free Bose gas}

Let us now  focus on spatial entanglement in free fields of indistinguishable particles, in particular for the Bosonic case. In contrast to the discrete spin models discussed so far, for second quantised fields even the term ``spatially separable''  is unclear and one has to establish a notion of what spatial entanglement can possibly mean. Second quantisation requires that one has to abstract from the particle picture and instead talk about the excitation of modes - in this case, spatial modes, see for instance the discussion in \cite{manifesto,06Heaney}. An easy way out is to simply find a reasonable description of what a spatially separable state shall be, without bothering about the mathematical description of spatial modes in terms of field operators  and complications that may arise on the way. Such a definition is already sufficient to establish entanglement witnesses whose expectation values are bounded for these separable states. This has been observed by Anders {\it et al.} in \cite{06Anders} and we summarise their results in the following. Nevertheless, in \cite{06Heaney} a first attempt has been made to define spatial modes constructively, in particular creation and annihilation operators which allow to talk about the entanglement in space instead of the usual momentum.

The argument in \cite{06Anders} goes as follows. A free Bosonic gas squeezed in a box of length $L$ is described by the second quantised Hamiltonian $H = \sum_k E_k a^{\dag}_k a_k$ with energies $E_k = {\hbar^2 \over 2m}\left({ k \pi \over L}\right)^2$ for $k =1, 2, ...$ in one dimension. The lowest possible energy which can be achieved by the system is the condensation energy $	E_{cond} = N {\hbar^2 \over 2m} \left({ \pi \over L}\right)^2$, for the case that all $N$ Bosons accumulate in the ground state. 
This is due to the Heisenberg uncertainty relation which requires a finite momentum for a finite spatial uncertainty. 
\begin{figure}[h]
  \begin{center}
   \scalebox{0.5}{\includegraphics{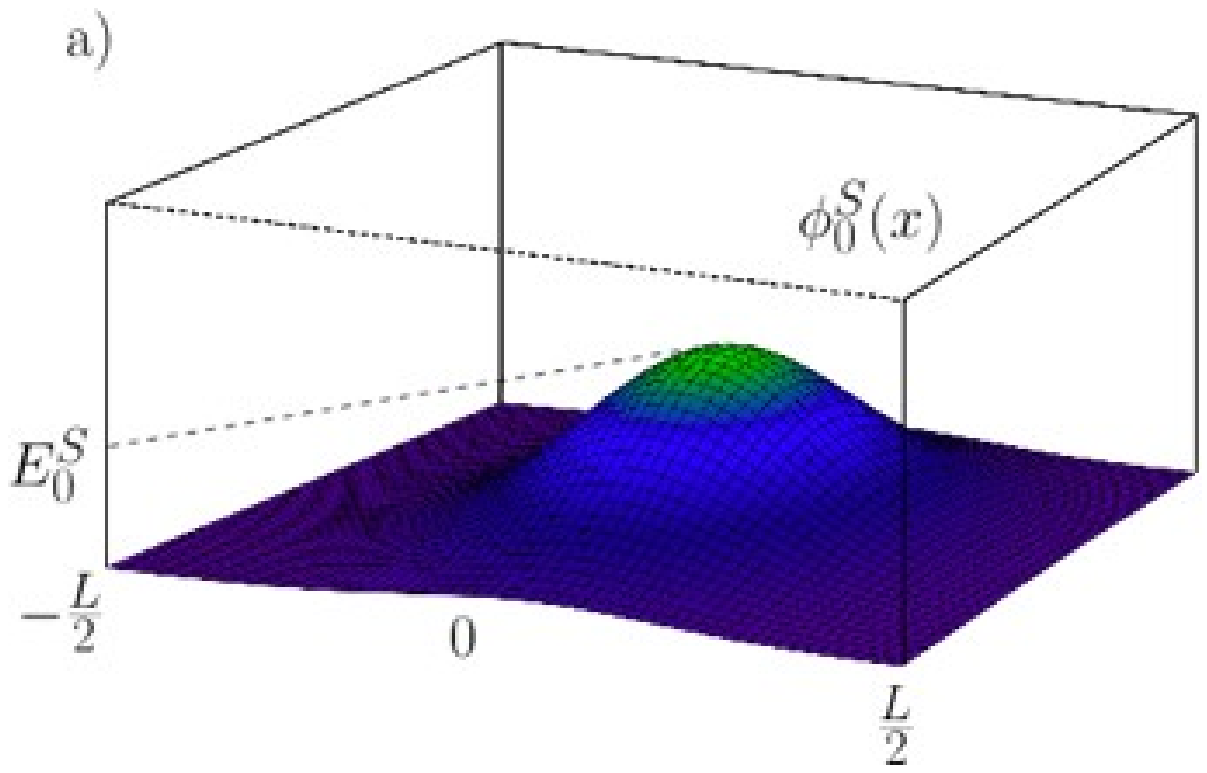}}
   \scalebox{0.5}{\includegraphics{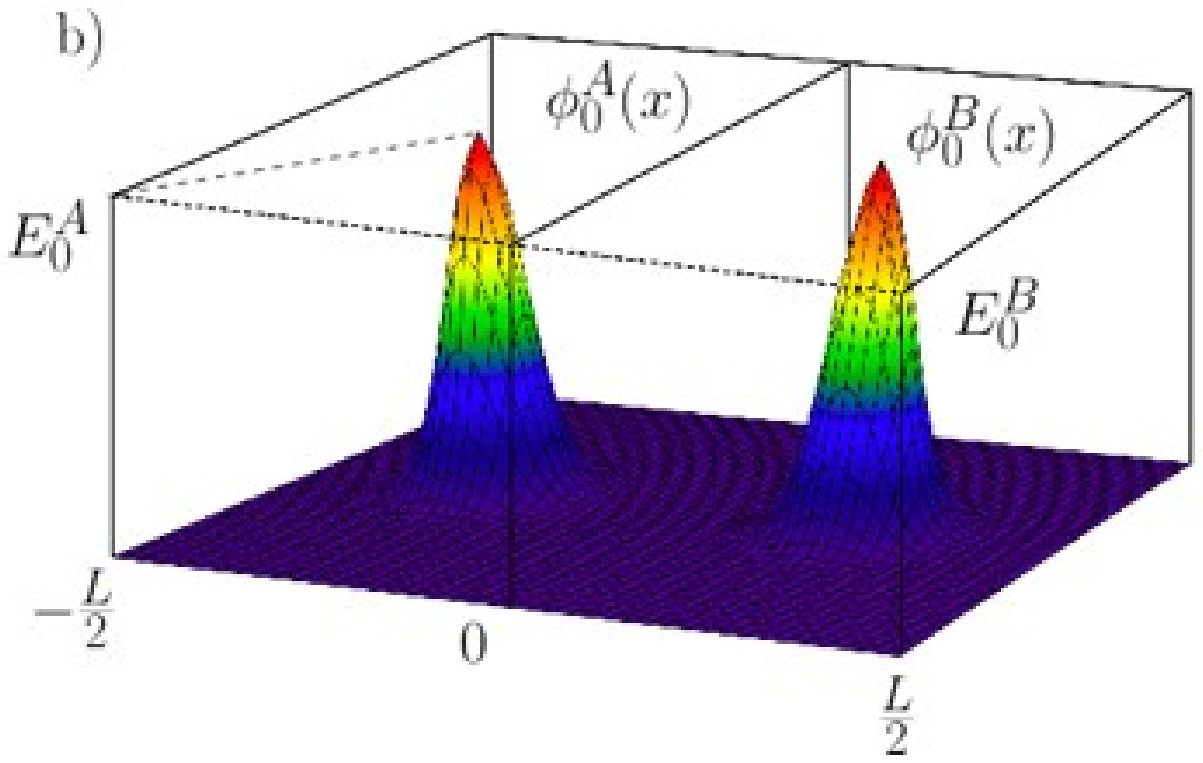}}
    \caption{\label{fig:bumps}   
    {\bf a)} The spatial distribution function for one Boson, $\phi_0^S (x)$, which  minimises the energy can spread over the whole box and has an associated energy of  $E_0^S = {\hbar^2 \over 2m} \left({ \pi \over L}\right)^2$. 
    {\bf b)}  A separable configuration with respect to the two regions $A$ and $B$ restricts the field to live in these regions separately. The respective energy-minimising distributions in $A$ and $B$ are $\phi_0^A (x)$ and $\phi_0^B (x)$ and require higher energies $E_0^A= E_0^B = 4 E_0^S $. This requirement makes separable configurations richer in energy and hotter in temperature.}
       \end{center}
 \end{figure}
To achieve a separable configuration in space, for example with respect to the two spatial regions $A = [0, L/2]$ and $B = (L/2,L]$, the field has to be squeezed to fit inside the new regions, see Figure \ref{fig:bumps}  for an illustration. The energy required for such a spatially separable configuration is then at least $E_{A|B} = N {\hbar^2 \pi^2 \over 2m} \left({2 \over L }\right)^2$, or higher.  Thus again we find, that the energy is a macroscopic entanglement witness:  It allows us to put a lower bound on the energy of separable configurations, or equivalently, those states which are detected having a lower energy must be entangled. 

The result was generalised to configurations of the field that are fully separable with respect to $M$ spatial regions in $d$ dimensions. By virtue of the thermodynamical relation between the energy and the temperature of the gas, $U = \tr [ \rho_T ~ H ] \propto T^{(d+2)/2}$, the authors derive a transition temperature in \cite{06Anders} where separability turns into entanglement,
\begin{equation}
	T_{trans} = {2 \pi \hbar^2 \over k_B m V^{2/d}} \, 
	\left({\pi N M^{2/d} \over 2 \zeta(1 + d/2)}\right)^{2/(2+d)},
\end{equation}
with $\zeta$ being Riemann's zeta-function and $V$ the volume of the confining box. If the number of regions, $M$, becomes equal to the number of particles $N$, than a situation is reached were each spatial region will on average be occupied by just one Boson. That implies that Bose-Einstein condensation (BEC) can only happen below $T_{trans} (M=N)$ which results in an upper estimate for the critical temperature of BEC, with $\rho=N/V$ for the particle density, it is 
\begin{equation}
	T_{crit} = {2 \pi \hbar^2 \, \rho^{2/d} \over k_B m} \, 
	\left({\pi \over 2 \zeta(1 + d/2)}\right)^{2/(2+d)},
\end{equation}
which looks very similar to the exact critical BEC temperature $T_{BEC}$ \cite{footnote1}. Indeed in 3D, $T_{crit}$ is only slightly higher than $T_{BEC}$. This is in good agreement with the fact that a BEC is entangled or spatially correlated.  However, even though the formulae look similar, there is a subtle difference.  In 1D and 2D BEC can in fact not occur, since the zeta-function $\zeta(d/2)$ in the denominator diverges for $d \le 2$ so that the critical BEC temperature drops to zero. Entanglement should, however, exist in any dimension, and that is indeed obeyed since  $\zeta(1+ d/2)$ is finite for any $d>0$.

Let us have another look at this issue. What is the physical reason for why there is no BEC in 1D and 2D?  BEC is the condensation of a finite fraction of the gas into the ground state (or any other state). However,  the density of states changes dramatically with the dimension, and we find that the fraction 
\begin{equation}
	{N \over N_0} \propto \int_0 {\dif^d p  \over e^{p^2/2m k_B T} -1}, 
\end{equation}
diverges  for $p \to 0$ as $1/p$  in 1D and logarithmically in 2D.  This divergence makes it impossible for BEC to occur; but not only that, such divergences are quite generally the mathematical reason why there is no conventional (long-range order) phase transition in low-dimensional systems.

\section{Phase transitions \& Order}\label{sec:PTs}

Criticality is a concept of central importance in solid state physics. Macroscopic objects, in the form of solids, liquids and gases, undergo a diverse range of phase transitions under variation of the temperature, or external field, or pressure for instance. Phenomena such as spontaneous magnetisation, or Bose-Einstein condensation of diluted atomic gases, are two notable examples of phase transitions in solids and gases respectively. According to Penrose and Onsager \cite{PO} they are manifested by the appearance of \emph{long-range order} (LRO), namely by the fact that distant constitutes of the system become, at the point of criticality, strongly correlated. 

When the two dimensional Ising model undergoes a phase transition at its critical temperature, this means that above this temperature, the spins were pretty much in a randomised, disordered state, where all directions are equally likely, but below that temperature all the spins align and point in some (randomly chosen) direction. This transition from a disordered to an ordered state is signified by \emph{spontaneous symmetry breaking} of the rotational invariance of the state of the Ising lattice. An additional parameter has to be introduced to specify which particular ``order'' was chosen by the system. The \emph{order parameter} in this case is the total magnetisation, which is finite below the point of criticality and becomes suddenly zero at and above this point. Symmetry breaking is, therefore, another indicator of phase transitions. 

\begin{figure}[t]
  	\begin{center}
   	\scalebox{0.4}{\includegraphics{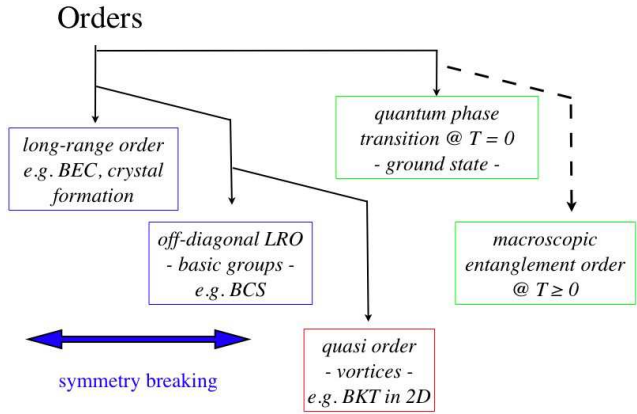}}
    	\caption{\label{fig:orders}  Overview of the different concepts of order: Long-range order (LRO) requires the coherence of all particles throughout the system whereas off-diagonal long-range order (ODLRO) assumes basic groups of particles which represent the relevant units in the ordered phase. Both long-range orders are a result of  a system undergoing spontaneous symmetry-breaking of a continuous symmetry and establishment of a non-zero order parameter.  In two dimensions the BKT theory predicts the binding and unbinding of vortices which creates local, short-range or quasi-order. The 2D phase transitions associated with the establishment of this kind of order do not break a symmetry and have no order parameter. A quantum phase transition  \cite{QPT} occurs when a system at $T=0$ abruptly changes its ground state under the variation of an external parameter. Similar to the overlap between two states of the field where a particle has been removed at two different positions, QPT can also be characterised in terms of the overlap function between two ground states \cite{Zanardi}. To see entanglement as an order concept by itself is a novel approach to characterise phase transitions in quantum systems which would unite all previous concepts under one roof.}
  	\end{center}
\end{figure}

Unfortunately, however, neither the appearance of long rang order, nor the phenomenon of symmetry breaking, are synonymous with phase transitions. There are instances where we do not have the establishment of long-range order, but a phase transition does take place, for instance the  Berezinskii-Kosterlitz-Thouless transition (BKT) \cite{BKT}. Therefore, while sufficient witnesses of phase transitions, the presence of long-range order and symmetry breaking are neither really necessary. Over the years, people have come up with ever new concepts of correlations and order to explain the occurrence of more exotic phase transitions. In the following we give a short overview of the zoo of the types of order serving as criticality indicators, see also Figure \ref{fig:orders}.

{\bf Long-range order (LRO)~}
The traditional way to characterise a phase as ordered is by requiring the establishment of long-range order \cite{PO} when the systems is brought into a critical parameter regime, i.e. by cooling below the critical temperature. Long-range order means, that the two-point correlation function for some relevant operator, $\hat a$, for two points in space does not vanish in the thermodynamical limit, even when the points are arbitrarily far apart,
\begin{equation}\label{eq:LRO}
	 \rho_1(x,y) : = \langle \hat a^{\dag} (x) \hat a (y)\rangle 
	 \underset{|x-y|\to \infty}{\longrightarrow} \mbox{const}.	
\end{equation}
The relevant operator $\hat a$ has to be identified for the phase transition under discussion. For instance, in the Ising model $\hat a (x)$ is the spin-operator $\sigma$ of the spin sitting at position $x$ and in a crystal it is the particle density at position $x$ in space. For fields, $\hat a(x)$ may become a field operator annihilating a particle at $x$. When LRO occurs,  any two points $x$ and $y$ in space become increasingly correlated and behave in a more and more coherent fashion, like for example spins slowly aligning themselves.  Also BEC is considered a long-range order transition, where the phase of the condensate wavefunction gets locked and all Bosons in the condensate behave as one single wave. 

{\bf Off-Diagonal long-range order (ODLRO)~}  
Already in 1962 Yang \cite{Yang} introduced a slightly weaker criterion of order for fields, for instance Fermionic fields, to accommodate the superconducting phase. Instead of looking at only two-point correlation functions for two modes $i$ and $j$, 
$\tr[ \hat a_j \rho  \hat a_i^{\dag}] = :\langle j| \rho_1 | i \rangle = \rho_1(i,j)$, 
Yang proposed to use the whole range of  generalised correlators, e.g. 
$ \tr[\hat a_k \hat a_l \rho  \hat a_i^{\dag} \hat a_j^{\dag}]=:\langle kl | \rho_2 | ij \rangle$ 
and so on, where the $\hat a_j (\hat a^{\dag}_j)$ are the annihilation (creation) operators for some single particle states $| j \rangle $. This implies that the original  $N$-particle state $\rho$ is reduced to $1, 2, ..., n$-particle states $\rho_1, \rho_2, ..., \rho_n$. Yang says ``The smallest $n$ for which ODLRO occurs, gives the collection of $n$ particles that, in some sense, forms a basic group exhibiting the long-range correlation." BEC is the simplest form of ODLRO, showing LRO already for $\rho_1$. 

To make this conclusion clear, let us suppose $\rho = |\psi \rangle \langle \psi|$ was a pure state. Then $\hat a_j |\psi \rangle$ is the state where one Boson has been removed at position $j$. If the Bosons form a condensate and are thus coherent, the state where instead a Boson has been removed at position $i$ should not differ by much from when it is removed at $j$. Thus the \emph{overlap} of these two cases, $ \langle \psi | \hat a^{\dag}_i \hat a_j |\psi \rangle = \tr[ \hat a^{\dag}_i \hat a_j \rho]=  \rho_1(i,j)$, becomes a constant and we do indeed find simple LRO according to Eq. \ref{eq:LRO}.

However, a superconductor does not have simple LRO. In fact we have to consider pairs of electrons, the \emph{Cooper pairs}, forming artificial Bosons. These Bosons can condense which shows up as LRO in the four-point-correlator $\rho_2$.  Since the condensed phase is made out of electrons and highly correlated throughout the medium, this phase can conduct current without any resistivity. This is the superconducting phase of the BCS theory \cite{BCS} from 1957.

{ \bf Quasi order ~}
 But also the concept of ODLRO was not enough to capture all known phase transitions. According to the famous Mermin-Wagner-Hohenberg Theorem \cite{Mermin, Hohenberg} no (OD)LRO can exist in one- and two-dimensional systems. However,  two-dimensional phase transitions were observed such as superfluidity and superconduction. An even more creative order concept, the \emph{quasi order} or sometimes \emph{topological order}, was introduced by Berezinskii and later by Kosterlitz and Thouless \cite{BKT}.  According to the BKT-theory, the transition to a quasi-ordered phase is driven by the pairing of  \emph{vortices} of opposite circulation, thereby creating short-ranged correlations without having the power to ever obtain (OD)LRO at a finite temperature. The achieved quasi- or short-range order decays polynomially fast for large separations. In terms of the two-point correlation functions this can be expressed as 
\begin{equation}
	\langle \hat a^{\dag}(x) \hat a (y)\rangle \underset{|x-y|\to \infty}{\longrightarrow} |x-y|^{- |\eta|},
\end{equation} 
where $|\eta|$ is a positive number.

\section{Entanglement \& Order}\label{sec:MEO}

In summary, phase transitions are intimately related to the idea of correlations and order between parts of the system. Trying to grasp what exactly this order is, that defines a new, macroscopically distinguishable phase, condensed matter physics has come up with a whole array of order concepts, tailored to fit the particular phase transition under discussion. However, macroscopic entanglement is a property of condensed matter systems which naturally grasps and quantifies correlations. So,  entanglement could be a good order concept itself, and its investigation may allow us to understand the underlying coupling mechanisms relevant for the abrupt change of the behaviour of a system at the critical point. \\

{\bf Entanglement~} Entanglement exists in all quantum systems, e.g. spin chains, free gases, lattices, ... and it seems that in all these systems entanglement can be seen as a macroscopic or thermodynamical property.  The essence of entanglement is that its existence implies correlations between parts of the system and entanglement has a huge capacity of capturing and quantifying any kind of correlations. 
Firstly, there is a choice of the degree of freedom which can be entangled, e.g. the spins in a magnetic chain, or the two possible states of a two-level atom, or the spatial modes of a field. Secondly, there are many different ways to arrange a grouping into parties. For example one can look at a bipartite split, e.g. two single spins in a chain or two blocks of spins, or at a multipartite split separating more than two parties.  These various groupings are in fact identical to Yang's basic groups which we encountered when discussing ODLRO. 

Unfortunately, multipartite entanglement is not well characterised and understood yet. However, it may be stimulating to use ideas from the area of condensed matter physics to find a characterisation, for example, Wen proposes \cite{Wen} to find a classification scheme for entanglement in analogy to the crystal classes of solids.

Of course, entanglement exists in all dimensions and in particular in low-dimensional systems thereby avoiding the problem of phase transitions in 2D. Entanglement thus appears to be a broad enough concept to grasp all the discussed ideas of order and correlation between distant parts of a quantum system \emph{at once}. The amount of entanglement may also be interpreted as an order parameter, becoming non-zero in the entanglement-ordered phase while zero in the separable, uncorrelated phase.

Similar to the notion of an ordered phase establishing abruptly at a critical temperature there is a transition from separability to entanglement at a certain transition temperature. 
We have seen that in 3D the critical temperature for BEC and the transition temperature for spatial entanglement are almost identical. Also,  Cooper pairs are entangled electron-pairs and therefore entanglement becomes an obvious ingredient for the occurrence of superconductivity, see also the  discussion in  \cite{04Vedral,Brandao}. Finally entanglement is an undisputed indicator of quantum phase transitions (QPT), that is transitions of the ground state at zero temperature, see for instance \cite{QPT,Zanardi}.  

All these properties indicate that entanglement may be an indicator for the occurrence of (some) phase transitions in quantum systems \cite{E=PT}. This is nice because entanglement is a fundamental property of quantum systems, and in particular a requirement for the occurrence of BEC and Cooper-pairing which indicates that we should maybe even see it as the cause  for such phase transitions. At the same time entanglement may also give us a natural explanation for phase transitions occurring in low-dimensional systems such as the short-range order Berezinskii-Kosterlitz-Thouless (BKT) transition in 2D. \\

{\bf Outlook~} So far so good, but why is all this exciting? The BCS theory of superconductivity has proven valuable for the interpretation of  the conventional superconduction phase transition, but it cannot explain the high critical temperatures (higher than 30K and up to 150K) of the \emph{unconventional} or \emph{high-$T_c$ superconductors} \cite{Bednorz}. High-$T_c$ super\-conductivity is a phase transition which happens in two dimensions  - the dimension where no (OD)LRO can occur according to the Mermin-Wagner-Hohenberg theorem.

\begin{figure}[b]
  	\begin{center}
   	\scalebox{0.5}{\includegraphics{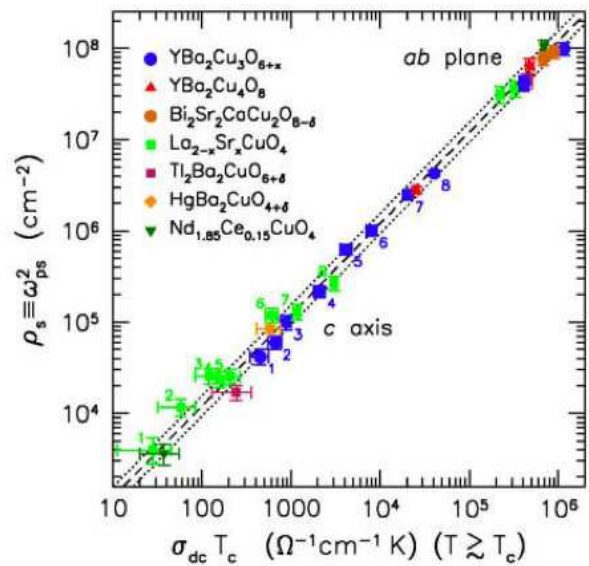}}
    	\caption{\label{fig:uemura-homes} Homes' law for the dependence of the critical temperature $T_c$ (horizontal axis) for seven different high-$T_c$ superconductors on the superconducting density $\rho_S$ (vertical axis). The linear law $\sigma T_c \propto  \rho_S$ becomes material independent when the  conductivity of the material $\sigma$ is included in the proportionality relation. This figure is taken from \cite{Homes}.}
       \end{center}
 \end{figure}

Nevertheless, experiments show that there is a phase transition and values of the critical temperatures of various materials have been investigated and published, e.g. in \cite{Uemura, Homes}. An empirical law, Homes' Law \cite{Homes, Zaanen}, was found which relates the critical temperature of a whole class of materials as directly proportional to the density of the superconducting phase $\rho_S$ times a proportionality factor which includes the electrical resistivity $R = 1/\sigma$ of the medium, $T_c \propto {1 \over \sigma} ~  \rho_S$, see the reprinted figure, Figure  \ref{fig:uemura-homes}, taken from \cite{Homes}. 

Remember that in two dimensions BEC was  impossible at any finite temperature, but we found the transition to an entangled and thereby correlated phase at the transition temperature $T_{crit} \propto \rho^{2/d}$ where $d$ is the dimension of the system. For $d=2$, this transition temperature $T_{crit}$ shows exactly the same form as Homes' law. Of course this may be coincidence, but more likely, a relation between high-$T_c$ superconductivity and the occurrence of entanglement exists as is the case for BEC in 3D and the BCS superconductivity. The name of the game is thus: Is a new theory explaining high-$T_c$ superconductivity based on entanglement? And how? Leaving this as the last contribution to this paper, let us summarise the discussion.

\section{Summary}\label{summary}

We saw that entanglement can be regarded as a macroscopic quantity in a variety of systems, such as Heisenberg chains and Bosonic gases. It can be tested by measuring thermodynamical quantities such as the energy or the magnetic susceptibility and it is possible to deduce (at least lower bounds on) the transition temperature for the transition from separability to entanglement. Further research clearly must address the question whether such macroscopic entanglement can exist at  high temperatures and how we can extract and use such natural entangled systems, for instance as quantum computers.  

In the second part of the paper we gave a short overview of the concepts of order occurring in the context of phase transitions. We have indicated that macroscopic entanglement may be a new concept of order, which could possibly unite all previous ideas such as long-range-, off-diagonal-long-range- and quasi-order. This would be a great simplification of the general theory of phase transitions and may lead to exciting predictions in particular for low-dimensional systems. Finally, macroscopic entanglement may be an essential ingredient for a new theory of high-$T_c$ superconductivity which remains a big challenge for us to solve.

\acknowledgements

The work summarised in this article is the result of our discussions and collaborations with many researchers who we wish to acknowledge here: M. Arnesen, S. Bose, \v{C}. Brukner, A. Zeilinger, D. Kaszlikowski, Ch.  Lunkes, T. Ohshima, L. Heaney, D. Markham, M. Murao and many others. J.A. is supported by the Gottlieb Daimler und Karl Benz-Stiftung. V.V. would like to thank the Engineering and Physical Sciences Research Council in UK and the European Union for financial support. This work was supported in part by the Singapore A*STAR Temasek Grant. No. 012-104-0040.


\begin{thebibliography}{99}

	\bibitem{06Heaney} 
	Heaney, L., Anders, J.  and Vedral, V.,  
		quant-ph/0607069 (2006).
	 
	\bibitem{ent-extract} 
	Kaszlikowski, D.  and Vedral, V.,  
		quant-ph/0606238 (2006);
	Terra Cunha, M.  O.  and Vedral, V.,  
		quant-ph/0607224 (2006)	.	 
     
	\bibitem{A} 	
	Nielsen, M. A., PhD-thesis, University of New Mexico (1998), 
		quant-ph/0011036; 
	
	\bibitem{B} 	
	Ghosh, S.,  Rosenbaum, T. F., Aeppli, G.  and Coppersmith, S. N.,  				
		{\it Nature} {\bf 425} (2003), 48; 
	Vedral, V.,  
		{\it Nature} {\bf 425} (2003), 28;
	Vedral, V.,  
		{\it New J. Phys.} {\bf 6} (2004), 102; 
	Wie\'sniak, M. , Vedral, V.  and Brukner, \v{C}.,   
		{\it New J. Phys.} {\bf 7} (2005), 258;	
	G\"uhne, O., T\'oth, G.  and Briegel, H. J.,  
		{\it New. J. Phys.} {\bf 7} (2005), 229;
	Vedral, V.,  
		{\it Nature} {\bf  439} (2006), 397;  
	and in particular \cite{Arnesen, Brukner, 06Brukner,06Anders,06Heaney,Markham}. 
	
	\bibitem{Arnesen} 
	Arnesen, M. C., Bose, S.  and Vedral, V.,  
		{\it \prl} {\bf 87} (2001), 017901.
	
	\bibitem{Brukner} 
	Brukner, \v{C}.,  and Vedral, V.,  
		quant-ph/0406040 (2004).
	
	\bibitem{06Brukner} 
	Brukner, \v{C}., Vedral,   V.,  and  Zeilinger, A.,  
		{\it \pra} {\bf 73} (2006), 012110.
	
	\bibitem{Berger} 
	Berger, L., Friedberg, S. A.  and Schriempf, J. T.,  
		{\it Phys. Rev.} {\bf 132} (1963), 1057.

	\bibitem{06Anders}  
	Anders, J., Kaszlikowski,   D., Lunkes,  Ch., Oshima,   T.  and Vedral, V.,  
		{\it New J. Phys.}  {\bf 8} (2006), 140.
	 
	 \bibitem{EWs} 
	 Horodecki, M., Horodecki,  P. and Horodecki, R.,  
	 	{\it Physics Letters A} \textbf{223} (1996), 1; 
	 Terhal, B. M.,  
	 	{\it J. Th. Comp. Sc.} \textbf{287(1)} (2002), 313.
	
	\bibitem{Markham} 
	Markham, D.,  Anders,  J., Vedral, V.  and Murao, M.,  
		quant-ph/0606103 (2006).
	
	\bibitem{QPT} 
	Vidal, G., Latorre,  J. I., Rico, E. and Kitaev, A.,  
		{\it \prl} {\bf 90} (2003), 227902;	
	Latorre,	J. I., Rico,  E. and Vidal, G.,  
		{\it Quant. Inform. Comput.} {\bf 4} (2004), 48;	
	Wu, L-A., Bandyopadhyay,   S. , Sarandy,  M. S.  and Lidar, D. A.,  
		{\it \pra} {\bf 72} (2005), 032309;
	Osterloh, A., Amico,  L., Falci, G. and Fazio, R.,  
		{\it Nature} {\bf 416} (2005), 608.

	\bibitem{Katsura} 
	Katsura, S.,  
		{\it Phys. Rev.} {\bf 127} (1962), 1508.
		
	\bibitem{manifesto} 
	Terra Cunha, M. O.,  Dunningham, J. A. and Vedral, V., 
		quant-ph/0606149 (2006).
	
	\bibitem{footnote1} The exact critical temperature for  BEC is 
	$T_{BEC} =  {2 \pi \hbar^2 \, \rho^{2/d} \over k_B m} \, \left({1\over  \zeta(d/2)}\right)^{2/d}$, see for instance, Pathria, R. K.,   {\it Statistical Mechanics}, Oxford: Butterworth-Heinemann,  1996. 

	\bibitem{PO} 
	Penrose, O.,  
		{\it Phil. Mag.} {\bf 42}  (1951), 1373;
	Penrose, O.  and Onsager, L.,  
		{\it Phys. Rev.} {\bf 104}  (1956), 576.

	\bibitem{BKT}  
	Berezinskii, V. L.,  
		{\it Zh. Eksp. Teor. Fiz.} {\bf 59} (1970), 907, 
		[{\it Sov. Phys. JEPT} {\bf 32} (1971), 493];
	Kosterlitz, J. M.  and Thouless, D. J.  
		{\it J. Phys.} {\bf  6}  (1973), 1181.
	
	\bibitem{Yang} 
	Yang, C. N.,  
		{\it Rev. Mod. Phys.} {\bf  34} (1962),  694.	
	
	\bibitem{BCS} 
	Bardeen, J., Cooper, L. N.  and Schrieffer, J. R.,  
		{\it Phys.  Rev.} {\bf 108} (1957), 1175.
	
	\bibitem{Mermin} 
	Mermin, N. D.  and Wagner, H.,   
		{\it \prl} {\bf 17} (1966), 1133.
	
	\bibitem{Hohenberg} 
	Hohenberg, P. C., 
		{\it Phys. Rev.} {\bf 158} (1967), 383.

	\bibitem{Zanardi} 
	Zanardi, P. and  Paunkovi\'c, N., 
		quant-ph/0512249 (2006).

	\bibitem{Wen} 
	Wen, X-G.,  
		{\it \prl} {\bf 90}  (2003), 016803.  
 
	\bibitem{04Vedral} 
	Vedral, V., 
	 	{\it New J. Phys.} {\bf 6} (2004), 102. 

	 \bibitem{Brandao} 
	 Brand\~{a}o, F. G. S. L.,  
	 	{\it New J. Phys.} {\bf 7} (2005), 24.
		
	\bibitem{E=PT}  
	Cavalcanti, D., Brand\~{a}o,  F. G. S. L. and Terra Cunha, M. O., 
	 	quant-ph/0510132 (2005).
	 	
	\bibitem{Bednorz} 
	Bednorz, J. G.  and M\"uller, K. A.,  
		{\it Z. Phys. B}. {\bf 64}  (1986), 189.

	\bibitem{Uemura} 
	Uemura, Y. J.,  {\it et al.}, 
		{\it \prl}  {\bf 62} (1989), 2317. 
	
	\bibitem{Homes} 
	Homes, C. C.,  {\it et al.}, 
		{\it Nature} {\bf 430} (2004), 539.
		
	\bibitem{Zaanen} 
	Zaanen, J.,  
		{\it Nature} {\bf 430} (2004), 512.	
	 	 
\end{thebibliography}
\end{document}